\documentstyle[aps,preprint,tighten,floats,epsf,rotate]{revtex}

\def\be{\begin{equation}}
\def\ee{\end{equation}}

\newcommand{\bpi}{\mbox{\boldmath $\pi$}}
\begin{document}
\draft


\preprint{\vbox{\it 
                        \null\hfill\rm    SI-TH-96-6, hep-ph/9612370}\\\\}
%
\title{Baryon number conservation and the formation of disoriented
chiral condensates}
\author{G. Holzwarth\thanks{%
e-mail: holzwarth@hrz.uni-siegen.d400.de}}
\address{Fachbereich Physik, Universit\"{a}t-GH-Siegen, 
D-57068 Siegen, Germany} 
%
%
\maketitle
\begin{abstract}
Conservation of baryon number as topological charge in effective chiral
field theories imposes a local constraint on the time evolution of field
configurations in the commonly used $O(4)$ model. Possible consequences for
the formation of chiral condensates after a quench from random initial 
configurations are discussed. 
It is argued that efficient dissipative terms are necessary
but not sufficient to unwind the randomly curled-up initial 
configuration, before collective motion towards the condensate can
grow. The existence of soliton stabilizing mechanisms 
will further prolong or prevent this process. 
\end{abstract}
%



\vspace{1 cm}

An attractive feature of the nonlinear sigma model and its 
extensions to higher chiral orders~\cite{ChPT} is the possibility to 
identify baryon number with the winding number~\cite{Skyrme} 
that characterizes the map of compactified coordinate space onto 
the $SU(2)$ manifold of the 
chiral fields $\Phi=(\sigma,\bpi)$, as long as they are constrained 
to the 3-sphere $\sigma^2+\bpi^2=f^2$. This concept has a
profound basis~\cite{Wi83} in the anomaly structure 
of the underlying QCD, and it has
led to numerous successful applications as efficient description for
meson-baryon systems without explicit fermionic fields.

In the $O(4)$ linear $\sigma$-model~\cite{GML}, on the other hand, the fields 
$(\sigma,\bpi)$ can freely explore the full 4-dimensional chiral space 
which allows for a convenient description of the possibility to restore the 
spontaneously broken chiral symmetry at finite temperatures~\cite{Baym},
to study critical fluctuations of the 
chiral condensate near the phase transition, or to follow the dynamical
formation of the condensate after a quench. 

Unfortunately, with the unconstrained embedding of the chiral field $\Phi$
into an $R^4$ manifold the concept of baryon number $B$ as
topological index is lost because the topological connectedness 
of this manifold is
trivial. Of course, in a parametrization like
\be
\label{Phi}
\Phi= R(x,t) U(x,t), ~~~~~~~~~~~ U \in SU(2)
\ee
one still can maintain the definition 
\be
\label{B}
B=\frac{1}{24 \pi^2} \int \;\epsilon^{i j k} tr L_i L_j L_k \;d^3x,~~~~~~~~~~
L_i = U^\dagger \partial_i U,
\ee
but this 'baryon number' $B$ changes its value whenever 
the field $R(x,t)$ as a function
of time passes through the origin $R$=0 for some value of $x$.

This is easily visualized in the $O(2)$ model in $1+1$ space-time
dimensions with two chiral fields, $\Phi=(\sigma,\pi)$,
\be
\label{sigpi}
\sigma(x,t)=r(x,t)\;\cos\phi(x,t),~~~~~~~~\pi(x,t)=r(x,t)\;\sin\phi(x,t).
\ee
All field configurations with boundary conditions 
\be
\label{bcsig}
\sigma(-\infty,t)=\sigma(+\infty,t)=f_\pi,~~~~~~~~~~~
\pi(-\infty,t)=\pi(+\infty,t)=0,
\ee
represent closed loops embedded in the $\sigma$-$\pi$ plane
which can be contracted continuously into the vacuum
configuration $\sigma\equiv f_\pi, \pi\equiv 0$, and the 'baryon number'
\be
\label{bnum}
B=\frac{1}{2\pi}\int_{-\infty}^{\infty} \frac{\partial}{\partial x}
\phi(x,t) dx
\ee
changes by one unit whenever the sling is pulled across the origin.

Soliton and vacuum configuration may be separated by a potential
barrier which could prevent for classical dynamics such baryon number
violating processes. 
Cohen~\cite{Cohen} has argued that the corresponding quantum 
mechanical tunneling amplitude vanishes
in tree approximation in the limit where the number of colors 
$N_c \to \infty$. But a possible suppression 
for finite $N_c$ through quantum corrections to our knowledge
has never been proven.
By suitable changes in the parameters of the model (e.g. by increasing the
temperature) the process can become even classically allowed.
This can be nicely demonstrated in the 1+1-dimensional
$O(2)$ model with a temperature dependent value of $f_\pi^2(T)$
because in this model the stabilization mechanism for solitons breaks
down already at temperatures which are much lower than the critical 
value for restoration of chiral symmetry. The lagrangian is 
\be
\label{GML}
{\cal L}(\Phi) = \frac{1}{2} \partial_\mu \Phi \cdot \partial^\mu \Phi
- {\cal V} (\Phi)
\ee
with the potential
\be
\label{pot} 
{\cal V} (\Phi) = \frac{\lambda}{4} (\Phi^2 - f^2)^2 - H \sigma \; .
\ee
In order to keep the minimum of ${\cal V}$ for finite $H$ at $\Phi^2 = f^2_\pi
(T)$ we define
\be
f^2 = f^2_\pi(T) - \frac{H}{\lambda f_\pi (T)} \; .
\ee
In 1+1 dimensions this model stabilizes solitons, as long as the
inequality
\be
\label{ineq}
\frac{H}{\lambda f_\pi^3} < 0.047
\ee
is satisfied. They correspond to bound trajectories of a classical
point particle moving in the potential $-{\cal V}$, starting from
the maximum of $-{\cal V}$ and returning to it after an infinite time.
These configurations sling 
around the hat of the potential 
${\cal V}$ close to (but inside of) the bottom of
the valley. In terms of $\sigma$- and $\pi$-masses
\be
\label{masses}
m_\pi^2=\frac{H}{f_\pi},~~~~~~~m_\sigma^2=2 \lambda f_\pi^2 + m_\pi^2
\ee
we find from (\ref{ineq}) approximately $m_\sigma > 6.5\; m_\pi$, as
stability condition.
The angular winding length $L_w\sim (\phi'(x=0))^{-1}$ 
(i.e. the inverse of the gradient of the angular field near the center
of the soliton) defines the typical spatial extent of the stable
'baryon'. For fixed ratio $m_\sigma/m_\pi$ it scales like
$(f_\pi^3/H)^{1/2}$. In the symmetry limit $H\to 0$ $\phi(x)$ approaches
$\pi(x/L+1)$, so in this limit the soliton completely occupies 
the spatial box of length $2L$. With increasing $L_w$ the radial bag
$r(x)$ gets very shallow, i.e. $r(x)$ stays very close to $f_\pi$
(cf. fig.1).

\begin{figure}[h]
\begin{center}
\leavevmode
\epsfysize=10truecm \vbox{\epsfbox{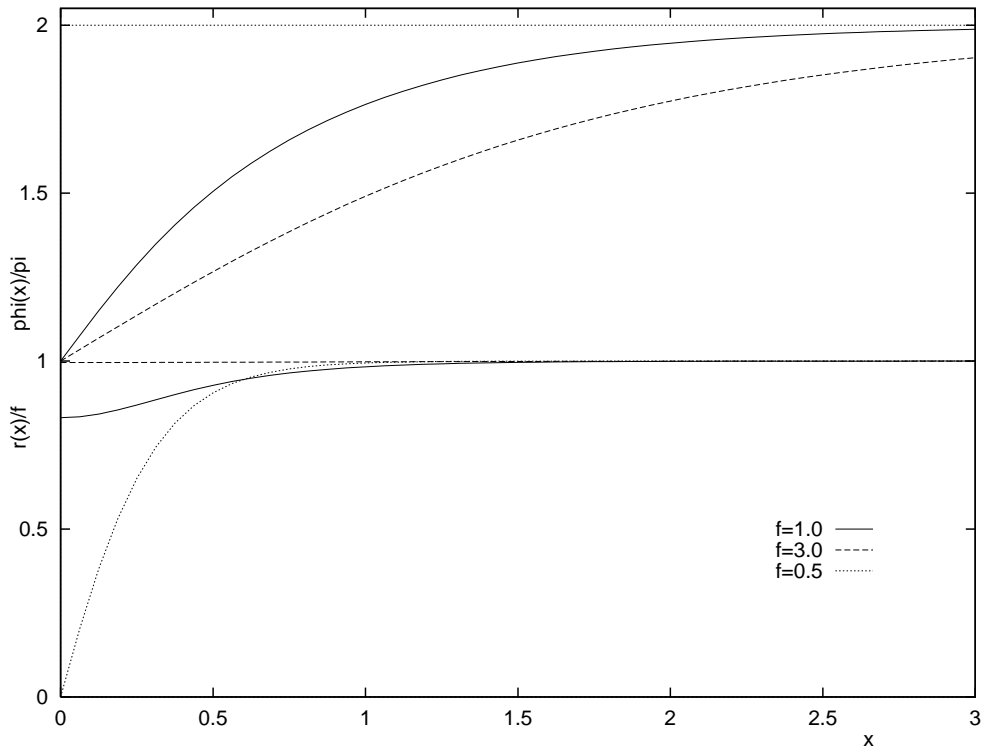}}
\end{center}
\caption{Soliton profiles in the $O(2)$ model ($m_\sigma=1000, m_\pi=138,
f_0=93$) for three different values
of $f_\pi=\alpha f_0$ ( full lines: $\alpha$=1.0; dashed
lines: $\alpha$=3.0;  dotted lines: $\alpha$=0.5 ). 
In the lower half the radial profiles
$r(x)/f$, in the upper half the corresponding
angular profiles $\phi(x)/\pi$ are shown.}
\label{fig1} 
\end{figure}

Suppose that $H$, $\lambda$ and $f_\pi$ satisfy condition (\ref{ineq})
for $T=0$  and suppose we have a 'baryon' represented by a stable 
trajectory with $B=1$. With increasing temperature $T$,
the inclusion of loop corrections into a renormalized ${\cal L}$
can lead to a decrease in $f_\pi^2(T)$ while
the coupling constant $\lambda$ and $H$ remain unrenormalized. 
This causes the soliton radius to decrease. At a
certain temperature the value of $H/(\lambda f_\pi^3)$ will exceed 
the critical stabilizing value (\ref{ineq}), 
the sling collapses to the vacuum point $\sigma\equiv f_\pi,\pi\equiv 0$, 
and the 'baryon' has disappeared. 

In the 3+1-dimensional $O(4)$-model soliton stabilization requires terms
of higher chiral order, like the fourth-order Skyrme term~\cite{Skyrme}
\be
\label{Skyrme}
{\cal L}^{(4)}=\frac{1}{32 e^2}\int tr[L_\mu,L_\nu]^2 d^3x.
\ee
Here the spatial extent of the angular chiral soliton profile is
determined by $(f_\pi(T) e)^{-1}$ where the 
Skyrme parameter $e$ remains essentially
unchanged with increasing temperature. Thus the Skyrme stabilization
mechanism works in the opposite way as compared to the stabilization
through the pion mass in the $O(2)$ model discussed above:
The angular winding length $L_w(T)\sim f_\pi(T)^{-1}$ 
as the characteristic length scale for stable field configurations now 
increases with increasing temperature. But as before, the
radial bag with increasing $L_w$ gets very shallow, 
i.e. $R(x)$ stays very close to $f_\pi(T)$ as $f_\pi(T)$ 
goes to zero. So for the static soliton the violation
of baryon number occurs together with the restoration of chiral
symmetry, but classical fluctuations $R(x,t)$ around the static 
configuration will cause fluctuations in baryon number 
already at lower temperatures.

Of course we do not want baryon number to be violated just by 
increasing the temperature. One possibility to avoid this 
would be to strictly stay within the nonlinear sigma model,
i.e. to impose the constraint $\sigma^2+\bpi^2=f_\pi^2(T)$ for all
configurations at all temperatures~\cite{Kapusta}. It is, however, not clear
how to describe in such a framework the 
system in the phase where chiral symmetry is restored.

A natural way to combine the convenient features of the linear 
$\sigma$-model with the topological advantage of the nonlinear $\sigma$-model
is to insist also in the $O(4)$ model on the angular and polar
nature of the chiral fields, i.e. on the parametrization (\ref{Phi}).
Because for $R$=0 the angular field $U$ is not defined the field 
configuration $\Phi$=0 then is excluded from the manifold
of possible chiral fields. This restores the nontrivial topological 
connectedness of the nonlinear $\sigma$-model. 
In our simple example of the 1+1 dimensional $O(2)$ model this 
means that we obtain topologically distinct classes of field 
configurations and consider the discrete manifold
\be
\label{nvac}
r=f_\pi,~~~~~\phi=n 2 \pi,  
\ee
($n$ integer) as a set of physically distinct degenerate vacua. 
The proper space in which the chiral fields then
live can be visualized either in polar representation $(r,\phi)$
as the half-plane $(~r>0,~~-\infty<\phi<\infty)$, or, in cartesian
representation $(\sigma,\pi)$ as multiple sheets tied together
in a multiple branch point at the origin.

If we now repeat the above considerations, starting with a stable 
baryon at $T=0$, increase the temperature, then at the critical 
value (\ref{ineq}) for $f_\pi(T)$ the sling now 
collapses to a narrow hairpin around the origin
which connects the vacua (\ref{nvac}) with $n=0$ and $n=1$,
$$
\phi(x)=2 \pi \Theta(x);
$$
\be
\label{hairpin}
r(|x|\to 0)=+0,~~~~~~~r(x\to\pm\infty)=f_\pi-Ae^{-m_\sigma|x|}.
\ee 
The stable nontrivial solution for $r(x)$ represents the 'bag' profile
of the remaining baryon with $B=1$. Although its angular winding length
$L_w$ has shrunk to zero, the radial bag still has a nonzero radius 
in coordinate space which
increases with temperature as long as the $\sigma$-mass decreases
(in this particular stabilization mechanism which is peculiar to
1 space dimension). In fig.1 soliton profiles are shown for some
values of $f_\pi^2(T)$ above and below the critical value. 
The stable soliton solutions above the critical
temperature do not exist if the point $\sigma$=$\pi$=0 is not excluded.

Of course, the exclusion of $R$=0 has severe consequences for the dynamical 
evolution of classical configurations.
Recently it has been suggested that the exponential growth of collective
amplitudes in the dynamical evolution of chiral field configurations
after a quench may lead to disoriented chiral condensates in
macroscopic regions of space~\cite{ans}-\cite{ggp}. 
It appears that these calculations have
been performed~\cite{rw} and confirmed~\cite{ggp} 
in the framework of the linear sigma $O(4)$ model~\cite{GML}
i.e. with chiral fields embedded in a simple $R^4$ internal space without
any constraints. It is an interesting question to ask whether and 
how the topological aspect discussed above may affect the results of
such calculations.

The typical difficulty appears already in the formulation of initial
conditions: Commonly they are taken as random Gaussian ensembles
centered around the origin $\sigma$=$\pi$=0 to reflect the chiral 
symmetry restored at temperatures higher than the chiral phase
transition temperature~\cite{rw}. In an ($r,\phi$) representation of the
chiral fields this would correspond to a uniform deviate in the 
angle $\phi$. It remains, however, ambiguous over how many multiples 
of $2\pi$ this uniform deviate extends. Chiral symmetry would require to
allow for all angles $-\infty<\phi<+\infty$ with equal probability,
which would imply that the initial field configuration could curl arbitrarily
often back and forth around the origin from one lattice point to the
next, with only the net winding number $n$ of the configuration inside
the box of size $2L$, 
i.e. the difference $2\pi n =\phi(L)-\phi(-L)$, fixed. 
Evidently, this is not a very meaningful choice for initial 
configurations. 

The high value of the temperature $T$ reflected in the initial
configuration implies a small 
correlation length $\xi_0\sim T^{-1}$ which characterizes 
the correlation function at $t=0$
\be
\label{corr}
<e^{i\phi(x_0+x,0)}e^{-i\phi(x_0,0)}>
=e^{-\frac{1}{2}<[\phi(x_0+x,0)-\phi(x_0,0)]^2>}=e^{-|x|/\xi_0}.
\ee
This shows that the mean square net winding increases linearly with
$x$. On a lattice with lattice constant $a$ this corresponds 
to a random walk with $|x|/a$ steps with a mean
square step size of $2 a/\xi_0$.
Identifying $\xi_0$ with the lattice constant we therefore sample 
initial configurations for the angles $\phi(x,0)$ by random walks with
Gaussian deviate for mean square stepsizes of order two, and fixed
value of total net winding $n$.
If $a<<2L$ these configurations again wind many
times back and forth around the origin, but most of the phase differences
between neighbouring lattice points are less than $2\pi$,
i.e. in other words, the local baryon density ($\sim \phi'(x,0)/(2\pi)$)
is small.    

Starting from such initial configurations, the singular point
at the origin causes a severe local barrier against the free
evolution of the classical trajectories: the growth of collective motion 
in radial direction towards 
the condensate can only take place (even for $n=0$) 
after unwinding the random multiple twists around the origin.
As long as the field is curled up around $r=0$, 
high- and low-$k^2$ Fourier components will have comparable power.
This can be verified in the 1+1
dimensional model discussed above, if the time evolution is performed
on a lattice in ($r,\phi$) representation.  
The actual value of the total winding number $n$ of the trajectory 
is of minor importance for the general features of the evolution during
this stage: as long as the field curls
many times back and forth around the origin, it is not 
essential at which multiple of $2\pi$ it ends.

The models discussed above do not even contain a term which
would drive the unwinding of the curled-up configuration. 
The evolution of the radial part $r(x)$ towards the condensate is
driven by the slope of the potential (\ref{pot}) whereas there is no
corresponding driving force to unwind the random large amplitude
fluctuations in the angle $\phi(x,0)$. This is easily demonstrated if 
(for the moment) we
keep the radial field fixed at some value $r(x)=r_0$ and then
separately consider the equation of motion for the angular part
\be
\label{singor}
\ddot\phi = \phi'' - \frac{H}{r_o} \sin \phi
\ee
which is of course the sine-Gordon equation and, as a typical wave 
equation conserves the mean amplitudes of randomly
fluctuating angles.

\begin{figure}[t]
\begin{center}
\leavevmode
\epsfysize=11truecm \vbox{\epsfbox{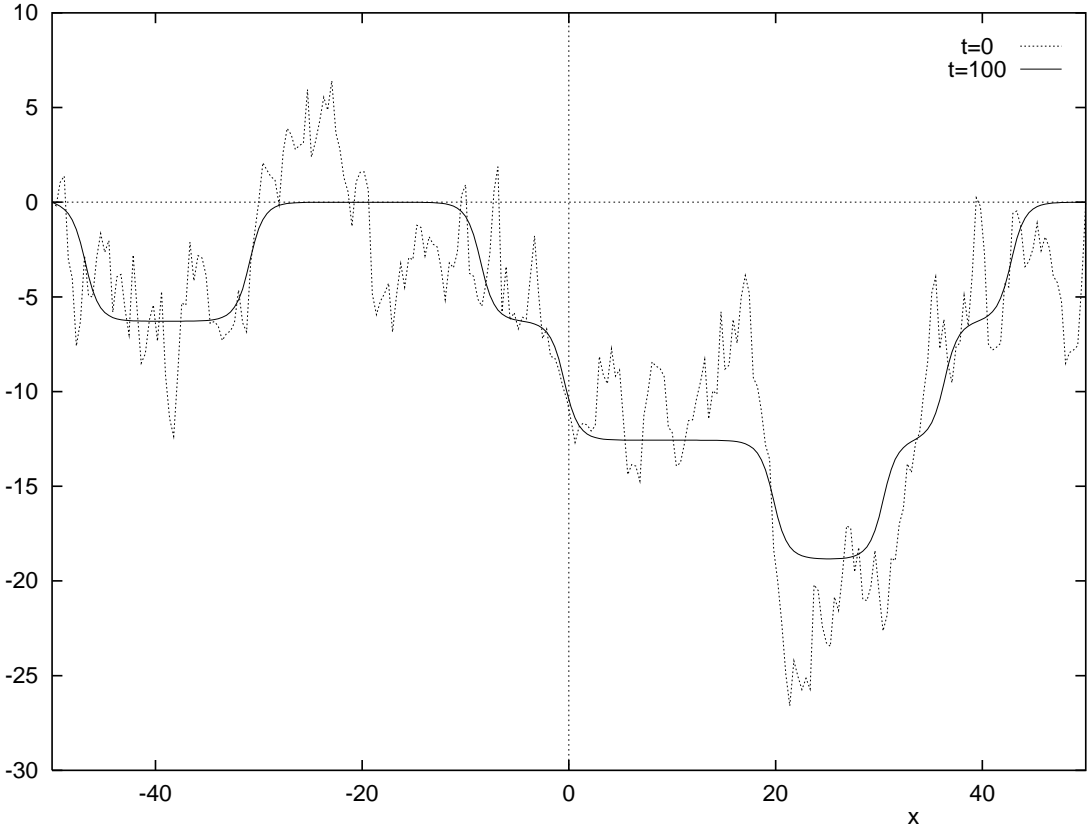}}
\end{center}
\caption{
The time evolution of the angular field $\phi$ in the $O(2)$
model from a random initial configuration $\phi(x,t=0)$ (dotted line) to a
quasistable soliton-antisoliton configuration $\phi(x,t=100)$ (full
line) for strong symmetry breaking ($m_\sigma=2000, m_\pi=1000, f_\pi=93$).}
\label{fig2} 
\end{figure}

Even the addition of a dissipative term $\dot\phi$ on the left hand 
side of (\ref{singor}) (which could be linked to
pion emission processes) to enforce the annihilation of large narrow
fluctuations does not necessarily lead to the unwinding of the
trajectories if the underlying lagrangian stabilizes solitons within the
box of length $2L$.
The dissipative term will relax the random fluctuations efficiently into a 
quasistable ensemble of $n_+$ solitons and $n_-$ antisolitons 
(with $n=n_+-n_-$) distributed 
randomly within the spatial box. Although the
configuration then is characterized by a phase correlation length of
order $L_w$ it generally will contain many sections winding back and
forth around the origin which
still prevent a definite orientation of the chiral field (even for $n=0$). 
Whether optimal
unwinding ($n_+=n$) can be achieved depends on the ratio of three
different length scales: the initial phase correlation length
$a$ ($\sim$ the lattice constant), the winding length $L_w$ of the stable
(anti)solitons, and the size $2L$ of the box. 
A necessary condition for unwinding is $a<<L_w$, because otherwise the 
initial random walk configuration will be close to some quasistable
soliton-antisoliton trajectory with a comparable large mean square
deviation $L/a$. If $a<<L_w<<L$ the quasistable trajectory can be very
different from the initial incoherent fluctuations but still may
contain many multiply curled-up sections. Further unwinding due to 
soliton-antisoliton annihilation will only take place if at some point
a soliton has sufficient overlap with an antisoliton.   
Only if $L_w$ is comparable or
larger than $2L/(n+2)$ then no stable soliton-antisoliton configurations 
fit into the box and the dissipative term can completely 
unwind all fluctuations. As we have
mentioned earlier this is always the case in the symmetry limit ($H=0$).

\begin{figure}[h]
\begin{center}
\leavevmode
\epsfysize=10truecm \vbox{\epsfbox{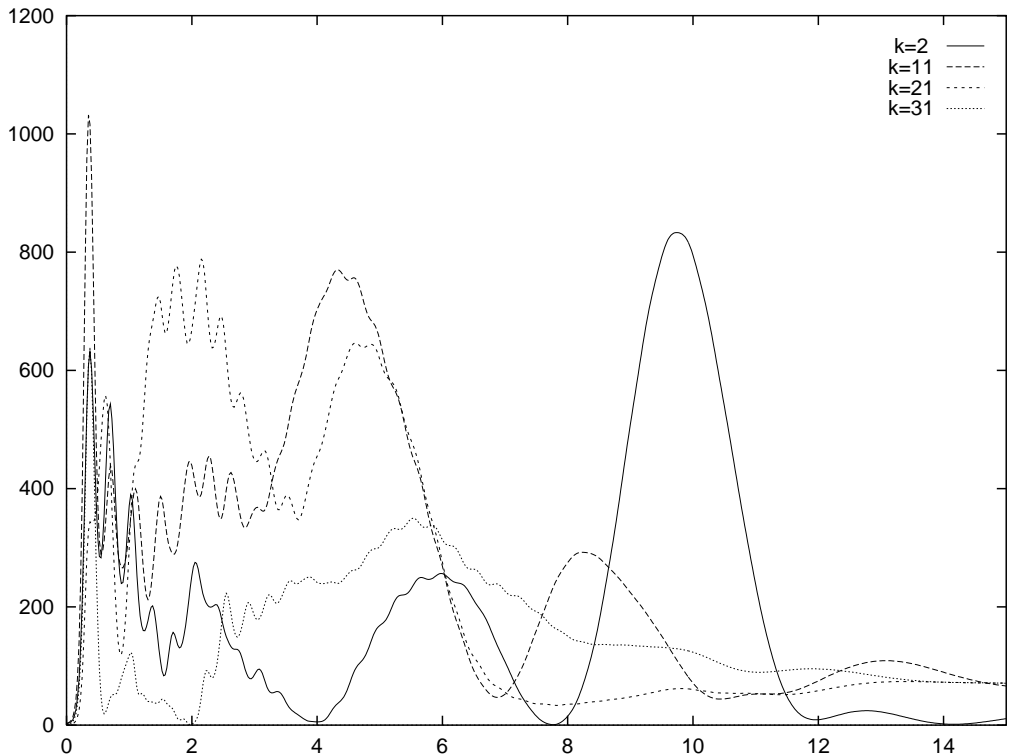}}
\end{center}
\caption{
The power of some of the 256 Fourier components $|\pi_k|^2$ 
(k=2,11,21,31) of the pion field $\pi(x,t)$ (3) (from $t=0$ up to $t=15$)
for the same evolution as shown in fig.2.}
\label{fig3} 
\end{figure}

Fig.2 shows a typical initial configuration for the angle $\phi(x,0)$
with $n=0$ for 256 lattice points. At the same time $r(x,0)>0$ is chosen
as narrow Gaussian deviate around $r=0$. Without dissipative terms 
$\phi(x,t)$ at much later times still looks like another random walk
with similar characteristics, i.e. it scatters over several multiples
of $2\pi$, while $r(x,t)$ then fluctuates around $r=1$. The
distribution of Fourier components $\phi_k$ remains essentially
unchanged. With dissipative terms and small symmetry breaking $H\to0$
the initial configuration of fig.2 slowly relaxes into the line 
$\phi \equiv 0$, while for strong symmetry breaking it evolves into the
quasistable soliton-antisoliton configuration shown in fig.2 at very
late time ($t=100$).

The power of some Fourier components $|\pi_k|^2$ (k=2,11,21,31)
of the pion field $\pi(x,t)$ (3) for the evolution of fig.2
is given in fig.3 up to $t=15$. At time $t\sim 10$ the final configuration
shown in fig.2 is essentially established. At earlier times  
there is no specific enhancement for the low-k
components.  

To our knowledge the ordering process in 
3+1 dimensional $O(4)$ models with soliton stabilizing dynamics 
has not yet been studied in detail. It might provide interesting
features of baryon-antibaryon formation through the cooling process
of the hot pion gas, as the different length scales change as functions
of the temperature. 
But the unexpected and anomalous exponents in the scaling behaviour
of 1+1-dimensional $O(2)$ spin systems after a quench
have recently been connected to topological textures in these 
models~\cite{bray}. 

It may appear as an unpleasant feature that the implementation of 
a global constraint $B=n$ should have such drastic consequences 
for the time evolution of the local field through the
requirement that a configuration should never move across the origin.
If it is true that the gross features of the time evolution are
insensitive to the actual value of $n$, 
one might be tempted to simply ignore this local constraint,
i.e. perform the calculation in a cartesian parametrization of the full
($\sigma$-$\bpi$) space of the $O(4)$ model, as it is conventionally
done, without distinguishing different vacua. 
This would imply that the total
baryon number will fluctuate during the time evolution of the field
as in a grand canonical ensemble with vanishing chemical potential.
It is, however, not clear under what conditions that may be 
a legitimate procedure.

To summarize, if we want to maintain the topological concept of 
baryon number conservation in 
effective chiral theories, we have to observe the angular nature of the
chiral field and allow for a multitude of physically distinct vacua. 
The global constraint imposed by the actual value of the 
baryon number may be unimportant. But the fact that the topological
charge is locally conserved has decisive influence on the dynamics
and may invalidade conclusions drawn from the time evolution
of chiral fields in the $O(4)$ model
without topological constraints.

\acknowledgements
The author would like to thank H. Walliser and H. B. Geyer for many
helpful discussions.


\begin{thebibliography}{99}

\bibitem{ChPT} S. Weinberg, {\em Physica} {\bf 96A}, 327 (1979);
J. Gasser and H. Leutwyler, {\em Ann. Phys. (N.Y.)}{\bf 158}, 142 (1984) 

\bibitem{Skyrme} T. H. R. Skyrme, {\em Prog. R. Soc.} {\bf A260}, 127 (1961);

\bibitem{Wi83} E. Witten, {\em Nucl. Phys.} {\bf B223}, 422 (1983), 
{\bf B223}, 433 (1983).

\bibitem{GML} M. Gell-Mann and M. Levy, {\em Nuovo Cimento}{\bf 16},
705 (1960)

\bibitem{Baym} G. Baym and G. Grinstein, {\em Phys. Rev.}{\bf D15}, 
2897 (1977)

\bibitem{Cohen} T. D. Cohen, {\em Phys. Rev.}{\bf D37}, 3344 (1988)

\bibitem{Kapusta} A. Bochkarev and J. Kapusta, {\em Phys. Rev.} {\bf
D54}, 4066 (1996).

\bibitem{ans} A.A. Anselm, {\em Phys. Lett.} {\bf B217}, 169 (1988); 
A.A. Anselm and M.G. Ryskin, {\em Phys.Lett.} {\bf B266}, 482 (1991).

\bibitem{bk1} J.P. Blaizot and A. Krzywicki, {\em Phys. Rev.} 
{\bf D46},246 (1992); {\em Phys. Rev.} {\bf
D50}, 442 (1994).

\bibitem{bj} J.D. Bjorken, {\em Int. J. Mod. Phys.} {\bf  A7}, 4189
(1992); {\em Acta Physica Polonica} {\bf B23}, 561 (1992). 

\bibitem{kt} K.L. Kowalski and C.C. Taylor, {\em Disoriented chiral 
condensate: A white paper for the full acceptance detector} 
CWRUTH-92-6 (1992), unpublished.

\bibitem{bjrev2} J.D. Bjorken, {\em Disoriented chiral condensate}, 
Proc. Workshop on Continuous advances in QCD, Minneapolis
 1994, and SLAC-PUB-6488 (1994). 

\bibitem{raj} K. Rajagopal, in{\em Quark-Gluon Plasma 2}, ed. R. Hwa
(World Scientific, 1995).

\bibitem{rw} K. Rajagopal and F. Wilczek, 
{\em Nucl. Phys.} {\bf B399},395 (1993), {\bf
B404},577 (1993). 

\bibitem{ggp} S. Gavin, A. Gocksch and R.D. Pisarski, {\em Phys.Rev.
Lett,}  {\bf 72}, 2143 (1994).

\bibitem{bray} A. D. Rutenberg and A. J. Bray, {\em Phys. Rev. Lett.}
{\bf 74}, 3836 (1995).

\end{thebibliography}
\end{document}